\documentclass[11pt,twoside]{article}

\usepackage{asp2014}

\aspSuppressVolSlug
\resetcounters

\begin{document}

\title{The PyKOALA python library: a multi-instrument package for IFS data reduction}


\author{Pablo~Corcho-Caballero,$^{1,2}$ Yago~Ascasibar,$^{3, 4}$ \'Angel~R.~L\'opez-S\'anchez,$^{5,6,2}$ 
Miguel~Gonz\'alez-Bol\'ivar,$^{7, 2, 6}$   Nuria~P.~F. Lorente,$^{7, 6}$ James Tocknell,$^{7, 6}$ Felipe~Jim\'enez-Ibarra,$^{7}$ Praveen Jayasuriya Daluwathumullagamage,$^{7}$ Gabriella Quattropani,$^{5,6,2}$, Matt Owers$^{5,6,2}$ and Gijs~A.~Verdoes-Kleijn$^{1}$}

\affil{$^1$Kapteyn Astronomical Institute, University of Groningen, Groningen, The Netherlands \email{p.corcho.caballero@rug.nl}}
\affil{$^2$The ARC Centre of Excellence for All Sky Astrophysics in 3 Dimensions}
\affil{$^3$Departamento de F\'isica Te\'orica, Universidad Aut\'onoma de Madrid, Madrid, Spain}
\affil{$^4$Centro de Investigaci\'{o}n Avanzada en F\'{i}sica Fundamental (CIAFF), Madrid 28049, Spain}
\affil{$^5$School of Mathematical and Physical Sciences, Macquarie University}
\affil{$^6$Astrophysics and Space Technologies Research Centre, Macquarie University}
\affil{$^7$Australian Astronomical Optics, Macquarie University, Sydney, NSW,
Australia}

\paperauthor{P.~Corcho-Caballero}{p.corcho.caballero@rug.nl}{0000-0001-6327-7080}{University of Groningen}{Kapteyn Astronomical Institute}{Groningen}{Groningen}{PO Box 800, 9700 AV Groningen}{The Netherlands}

\paperauthor{Y.~Ascasibar}{yago.ascasibar@uam.es}{0000-0003-1577-2479}{Universidad Aut\'onoma de Madrid}{Departamento de F\'isica Te\'orica}{Madrid}{Madrid}{E-28049}{Spain}

\paperauthor{\'A.~R.~L\'opez-S\'anchez}{angel.lopez-sanchez@mq.edu.au}{0000-0001-8083-8046}{Macquarie University}{School of Mathematical and Physical Sciences}{Sydney}{NSW}{2113}{Australia}

\paperauthor{M.~Gonz\'alez-Bol\'ivar}{miguel.gonzalez-bolivar@mq.edu.au}{0000-0002-5939-9269}{Macquarie University}{Australian Astronomical Optics (AAO)}{Sydney}{NSW}{2113}{Australia}

\paperauthor{N.~P.~F.~Lorente}{nuria.lorente@mq.edu.au}{0000-0003-0450-4807}{Macquarie University}{Australian Astronomical Optics (AAO)}{Sydney}{NSW}{2113}{Australia}

\paperauthor{J.~Tocknell}{james.tocknell@mq.edu.au}{0000-0001-6637-6922}{Macquarie University}{Australian Astronomical Optics (AAO)}{Sydney}{NSW}{2113}{Australia}

\paperauthor{F.~Jim\'enez-Ibarra}{}{0000-0002-4634-1076}{Macquarie University}{Australian Astronomical Optics (AAO)}{Sydney}{NSW}{2113}{Australia}

\paperauthor{P.~J.~Daluwathumullagamage}{}{}{Macquarie University}{Australian Astronomical Optics (AAO)}{Sydney}{NSW}{2113}{Australia}

\paperauthor{G.~Quattropani}{gabriella.quattropani@hdr.mq.edu.au}{}{Macquarie University}{School of Mathematical and Physical Sciences}{Sydney}{NSW}{2113}{Australia}

\paperauthor{M.~Owers}{matt.owers@mq.edu.au}{0000-0002-2879-1663}{Macquarie University}{School of Mathematical and Physical Sciences}{Sydney}{NSW}{2113}{Australia}

\paperauthor{G.A.~Verdoes-Kleijn}{g.a.verdoes.kleijn@rug.nl}{0000-0001-5803-2580}{University of Groningen}{Kapteyn Astronomical Institute}{Groningen}{Groningen}{PO Box 800, 9700 AV Groningen}{The Netherlands}


\begin{abstract}
PyKOALA is an innovative Python-based library designed to provide a robust and flexible framework for Integral Field Spectroscopy (IFS) data reduction.
By addressing the complexities of transforming raw measurements into scientifically valuable spectra, PyKOALA simplifies the data reduction pipeline while remaining instrument-agnostic and user-friendly.
This proceeding outlines the challenges of IFS data reduction, PyKOALA's architecture, and its applications to observations by the KOALA+AAOmega instruments at the Anglo-Australian Telescope.

\end{abstract}



\section{Introduction}

Integral Field Spectroscopy (IFS) has emerged as a transformative technique in astronomical research, enabling simultaneous spatial and spectral data acquisition across an extended field of view 
via creating complete three-dimensional view of astronomical objects.
This capability has revolutionized our understanding of the Universe by facilitating studies of complex astrophysical phenomena, especially in the field of galaxy formation and evolution \citep[e.g.,][]{Sanchez+21}.

Over the past two decades, advances in instrumentation and data processing have positioned IFS at the forefront of observational astronomy.
Flagship IFS surveys and instruments like the Calar Alto Legacy Integral Field Area (CALIFA) survey \citep{Sanchez+12} with PMAS/PPAK \citep{Kelz+06}, the many observational programmes carried out by the Multi-Unit Spectroscopic Explorer \citep[MUSE,][]{Bacon+10} at the Very Large Telescope, the SAMI Galaxy Survey \citep{Croom+12},
or the 
MaNGA survey \citep{Drory+15, Bundy+15} have demonstrated the power of IFS in delivering unprecedented insights into the nature of galaxies and their environments.

The wealth of data generated by IFS requires sophisticated reduction pipelines to transform raw measurements into scientifically useful information \citep[e.g.,][]{Weilbacher+20}.
This process encompasses multiple steps: correcting instrumental and atmospheric effects, calibrating fluxes, building 3D datacubes.
Each instrument poses specific requirements for these tasks, often addressed by custom software and expertise.
The Kilofibre Optical AAT Lenslet Array instrument \citep[KOALA,][]{Ellis+12}, mounted at the Anglo-Australian Telescope (AAT), is an Integral Field Unit (IFU) feeding the double-beam AAOmega spectrograph \citep{Smith+04}  
that perfectly exemplifies these challenges.
KOALA+AAOmega versatility (multiple gratings, central wavelengths, field of view) adds 
layers of complexity to the data reduction process.

The diversity and complexity of IFS instruments sets a significant barrier to the scientific exploitation of these data.
To mitigate this issue, the PyKOALA project started as a bespoke data reduction package for KOALA+AAOmega, but now it has grown to become a comprehensive, instrument-agnostic solution for IFS data reduction.
This paper introduces PyKOALA, detailing its philosophy, structure, and current features.

\section{PyKOALA}

The core design principles of PyKOALA\footnote{\url{https://pykoala.readthedocs.io/}} are threefold: to provide a comprehensive framework adaptable to any IFS instrument; to establish a modular and extensible structure enabling customized workflows; and to enhance user accessibility through detailed documentation and tutorials.

Astronomical data in PyKOALA is managed through \texttt{DataContainer} objects, designed to offer a uniform and efficient means of handling observations from various instruments. PyKOALA's API supports specific representations for two common formats widely used in IFS:
Row-Stacked Spectra (\texttt{RSS}), that represents the spectra collected by each individual fibre in the IFU after tracing and extraction from the raw exposures, and datacubes (\texttt{Cube}), a 3D grid representation of spectra, sampled along one spectral and two spatial dimensions.

\articlefigure[width=0.85\linewidth]{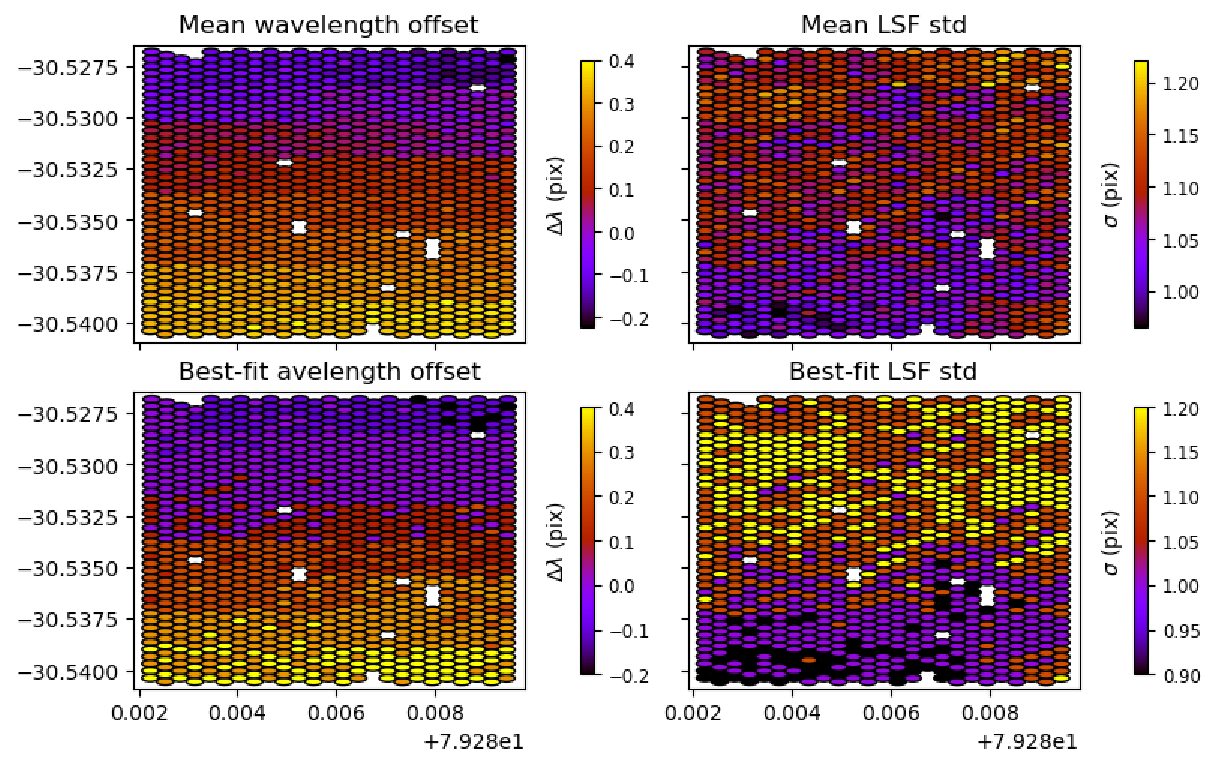}{wave_offset}{Wavelength offset (left) and LSF standard deviation (right) for an RSS twilight exposure. Top and bottom rows display the results obtained from a weighted-average and a maximum-likelihood approach, respectively.}

Since PyKOALA's development has been significantly driven by the requirements of KOALA+AAOmega, its initial focus has been on processing wavelength-calibrated RSS and datacubes, although the algorithmic implementation has been tested with other instruments (Hector and WEAVE).
For KOALA+AAOmega data, the library includes dedicated methods for reading the RSS files produced by the 2dfdr pipeline \citep{aao_team+15}.
For data from other instruments, users can integrate PyKOALA by providing an interface (e.g., a wrapper function) to extract necessary information -- such as fibre spectra, wavelength, and variance arrays -- from the instrument's data files.
These inputs can then be used to instantiate an RSS class, leveraging the existing framework that provides an efficient integration.

Reduction steps in PyKOALA, such as atmospheric extinction correction, sky subtraction, or astrometry adjustments are implemented via specialized \texttt{Correction} classes.
This modular approach provides a flexible and intuitive framework, allowing users to easily define and customize data reduction sequences.
PyKOALA currently supports the following corrections: atmospheric effects such as extinction, sky emission, and telluric absorption; astrometry offset corrections; wavelength offset corrections; and flux calibration.

As an example, Fig.~\ref{wave_offset} illustrates the fibre-to-fibre wavelength correction and Line Spread Function (LSF) estimation of a single twilight exposure, performed using cross-correlation with a reference solar spectrum. This step is critical for achieving sub-pixel wavelength calibration accuracy, which is essential for subsequent reduction steps such as precise sky subtraction.

PyKOALA also supports the interpolation of individual RSS exposures into 3D datacubes, offering users the flexibility to choose between inverse-distance weighted methods and alternative techniques, such as drizzling \citep{Fruchter+02}.
Fig.~\ref{cubing} showcases three different interpolation approaches applied to the same set of RSS exposures, resulting in slightly distinct cube reconstructions.
The choice of interpolation method depends on the nature of the data and specific scientific objectives.

\articlefigure{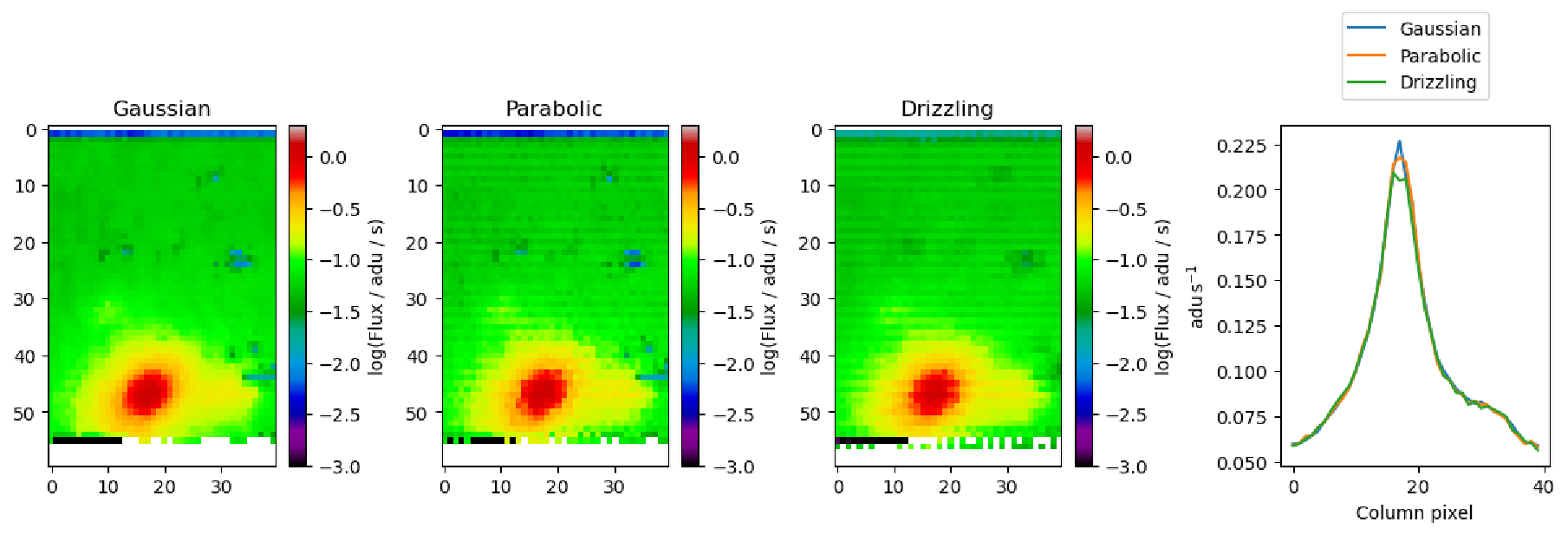}{cubing}{Example combining a set of RSS data into a datacube using a Gaussian (left), Parabolic (middle) and Drizzling (right) interpolation kernel. The rightmost panel shows the light profile of each datacube, collapsing the first spatial dimension.}

\section{Conclusions and ongoing work}

PyKOALA advances the field of IFS data reduction, by providing a robust, flexible, and instrument-agnostic framework that addresses the complexities of modern data.
This is achieved via a modular structure, intuitive API, and comprehensive documentation.

One of the library's key strengths is its adaptability, allowing it to support multiple instruments through the implementation of custom wrappers, while maintaining a consistent and efficient workflow.
By leveraging a modular architecture with specialized Correction classes, PyKOALA provides users the freedom to design customized reduction pipelines tailored to their specific scientific needs.
The library's development has been closely aligned with the needs of the KOALA instrument, for which it will serve as the primary data reduction tool together with 2dfdr.
At the same time, PyKOALA's design includes future extensions to support additional instruments and enhanced compatibility with widely-used astronomical software ecosystems such as Astropy.

Initial testing of PyKOALA has demonstrated its capability to achieve precise corrections, including sub-pixel wavelength calibration, effective sky subtraction, and accurate flux calibration.
Moreover, its interpolation techniques for generating 3D datacubes provide flexibility for various scientific applications, ensuring that the resulting data products meet diverse research goals.

We plan to release a stable version of PyKOALA in Q1 2025, including introductory tutorials and documentation.
The project is an open collaboration, and welcomes contributions from the broader community.

\bibliography{C607_v1}  


\end{document}